\documentclass{article}

\PassOptionsToPackage{numbers,sort&compress}{natbib}
\usepackage[preprint]{pai26}
\usepackage{todonotes} 

\workshoptitle{2026 Conference on Physics and AI (PAI26)}

\usepackage[utf8]{inputenc} 
\usepackage[T1]{fontenc}    
\usepackage{hyperref}       
\usepackage{url}            
\usepackage{booktabs}       
\usepackage{amsfonts}       
\usepackage{nicefrac}       
\usepackage{microtype}      
\usepackage{xcolor}         
\usepackage{amsmath} 

\usepackage{doi}
\usepackage[most]{tcolorbox}
\usepackage{xcolor}
\usepackage{makecell}

\title{Neutrino Fingerprints: Image-Based Encodings of IceCube Events for CNN Direction Reconstruction}

\author{%
  Brecht Verbeken$^{1,2}$\thanks{Correspondence to \texttt{brecht.verbeken@vub.be}. \\The code is publicly available at \url{https://github.com/Integrated-Intelligence-Lab/neutrino-detection-experiment}}, Floriano Tori$^{1,2}$, Vincent Ginis$^{1,2,3}$ \\
  $^{1}$Data Analytics Lab, Vrije Universiteit Brussel, Pleinlaan 5, 1050 Brussels, Belgium \\
  $^{2}$imec-SMIT, Vrije Universiteit Brussel, Pleinlaan 9, 1050 Brussels, Belgium \\
  $^{3}$School of Engineering and Applied Sciences, Harvard University, Cambridge, Massachusetts 02138, USA
}

\begin{document}

\maketitle

\begin{abstract}
Reconstructing the direction of incoming neutrinos in the IceCube Neutrino Observatory is an important problem in astrophysics. The public IceCube--Neutrinos in Deep Ice Kaggle competition provided 140 million simulated events to benchmark reconstruction techniques. To address this challenge from a novel perspective we introduce \emph{neutrino fingerprints} compact $72 \times 72 \times 3$ images in which each pixel represents a single detector, with pulse timing and charge statistics encoded as color channels.
This representation transforms sparse, irregular pulse data into dense images suitable for convolutional processing. Our ResNet18 model achieves a mean angular error of $1.10$ rad, indicating that convolutional networks trained on fingerprints rival more complex architectures while offering an effective, interpretable baseline for IceCube event reconstruction.
\end{abstract}

\section{Introduction}
Direction reconstruction in IceCube is a longstanding inverse problem: Cherenkov photons emitted by charged secondaries propagate through kilometre--scale glacial ice and are detected by a sparse three--dimensional array of photomultiplier detectors. Traditional reconstruction algorithms rely on analytic approximations to the photon transport likelihood and can reach sub--degree angular resolution, but their high computational cost makes them impractical for large datasets~\cite{abbasi2022low}. Neural networks
can dampen this inference cost by learning a mapping from raw detector responses to direction vectors.  Early studies used convolutional neural networks (CNNs) for GeV and TeV--scale events\cite{abbasi2021convolutional,yu2021direction}, while recent work has shifted toward graph neural networks and transformers, culminating in a public Kaggle competition and numerous high--performing solutions \cite{bukhari2024icecube, choma2018graph}. In this work we revisit CNNs by introducing \emph{neutrino fingerprints}, image encodings in which each detector is represented by a pixel and its pulse statistics by color channels, and evaluate how competitive such a simple baseline can be. Beyond direction reconstruction, a parallel line of work has focused on classifying event topologies (track--like vs.\ cascade--like) \cite{choma2018graph, abbasi2021convolutional, kronmueller2019inproceedings}.

IceCube has demonstrated that convolutional architectures such as InceptionResNet can deliver competitive performance for topology classification
when events are embedded on a regular grid~\cite{kronmueller2019inproceedings}. Our study is complementary: instead of event type classification, we we return to CNNs for directional reconstruction by introducing \emph{neutrino fingerprints}, compact image encodings in which each detector is represented by a pixel and its pulse statistics by color channels.
 Before presenting our approach, we summarise the IceCube detector and the competition dataset.

\section{IceCube detector and simulated dataset}

\subsection{Detector geometry}
The IceCube Neutrino Observatory equips a cubic kilometre of Antarctic ice with 5\,160 photomultiplier detectors deployed on 86 vertical strings between depths of 1.45 and 2.45\,km. The main array comprises 78 strings with 125\,m horizontal and 17\,m vertical spacing, while the 8--string \emph{DeepCore} subarray uses 70\,m and 7\,m spacing to extend sensitivity down to $\mathcal{O}(10\,\mathrm{GeV})$ neutrinos~\cite{icecubecollaboration2025fastlowenergyreconstruction}. The resulting detector layout is sparse and irregularly distributed in three dimensions, making it challenging for learning algorithms that assume regular grids. A right--handed coordinate system is used with the origin near the detector centre, $+z$ upward to the surface, $+x$ east, and $+y$ north~\cite{abbasi2009icecube}. 

\subsection{Data characteristics and pulse features}
Each of the 140 million IceCube events is recorded as a variable--length sequence of pulses produced when Cherenkov photons reach the photomultiplier detectors. For the Kaggle data, every pulse is described by four quantities: the time within the event window (ns), the integer \texttt{sensor\_id} of the detector that registered it, the measured charge (photoelectrons), and a Boolean \texttt{auxiliary} flag indicating reduced--quality readout~\cite{bukhari2024icecube}. A separate geometry file provides the fixed $(x,y,z)$ positions of all 5\,160 detectors, keyed by \texttt{sensor\_id}~\cite{abbasi2009icecube}. Thus the effective representation of an event is an $n_{\text{pulses}}\!\times\!(\texttt{time},\texttt{sensor\_id},\texttt{charge},\texttt{aux})$ table joined with static detector coordinates. 

In our work we restrict to the three features \texttt{time}, \texttt{sensor\_id}, and \texttt{charge}. The \texttt{auxiliary} flag, which marks pulses more likely to be noise, is ignored, although incorporating it might improve discrimination. Pulse multiplicities are highly variable, from only a few to more than $10^5$, which complicates direct use in fixed--size models and motivates our image--based fingerprint encoding.

The simulated events span energies from 100\,GeV to 100\,PeV and include all neutrino flavours and interaction types, giving rise to both track--like and cascade--like topologies \cite{denton2018invisible}. These events were grouped into 660 random sub-samples, so-called "data batches", each containing the detector response from 200.000 events.
 Metadata files specify for each event its \texttt{event\_id}, \texttt{batch\_id}, pulse index ranges within the batch table, and (for training events) the true azimuth and zenith angles of the incoming particle. No additional pulse cleaning is applied, so the dataset reflects realistic detector noise and medium uncertainties.

\section{Event Representation and Model Setup}
Our core idea is to cast the irregular, sparse detector response into a fixed--size image representation, enabling the direct use of convolutional networks. Each event is distilled into a compact \emph{neutrino fingerprint}: a $72 \times 72 \times 3$ image where every pixel corresponds to a single detector. The image channels store simple yet informative pulse statistics. In this section we describe the encoding, network, and training setup.

\subsection{Constructing the Fingerprint}
The mapping from detector index $d \in \{0,\dots,5159\}$ to image coordinates is given by a deterministic row--major ordering,
\[
x(d) = d \bmod 72, \qquad y(d) = \left\lfloor \tfrac{d}{72} \right\rfloor .
\]

This row-major rasterization is a naive 3D-to-2D locality proxy: adjacent detector indices remain adjacent pixels, but some physical neighbors, e.g. across strings or index-block boundaries, are separated.
This covers the full array, leaving 24 unused pixels that are filled with a constant white background value, RGB=(255,255,255), identical across all events. A pixel and a detector are thus synonymous in our encoding.

\begin{figure}[h!]
    \centering
    \includegraphics[width=1.0\textwidth]{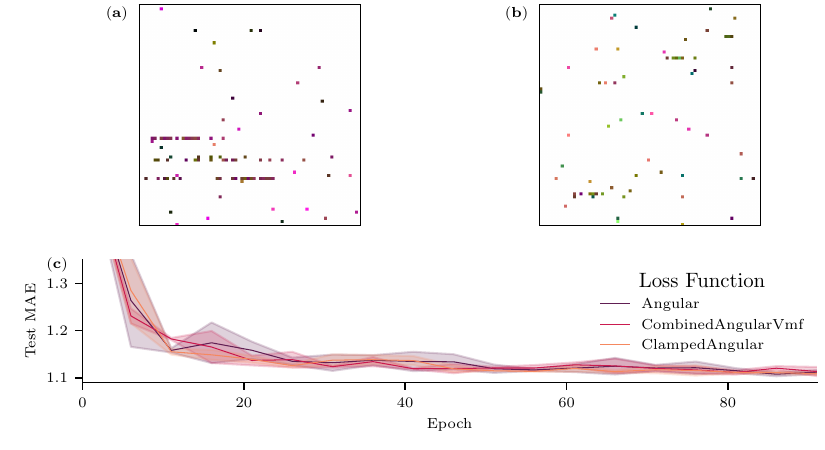}
    \caption{\textbf{We show that 2D images of neutrino events, using a convolution based architectures, can be used to reconstruct the directionality of an event.} (a.) A cascade type event transformed using our fingerprint method. (b.) Standard event transformed (c.) Test Mean Angular Error evolution during training for different losses averaged over three seeds. }
    \label{fig: main_figure_1}
\end{figure}

From the variable--length pulse sequence of each event we summarize every detector $d$ by two scalars: a total charge $Q_d$ and an earliest arrival time $T_d$. To stabilize intensities across events with very different multiplicities and energies, we normalize each variable over the active detector set of the event $\mathcal{D}$:
\[
\mathrm{Norm}(v_d) \;=\; 255 \cdot \frac{v_d - \min_{d'\in \mathcal{D}} v_{d'}}{\max_{d'\in \mathcal{D}} v_{d'} - \min_{d'\in \mathcal{D}} v_{d'}} .
\]
This per--event rescaling forces the network to attend to relative spatial--temporal patterns rather than absolute scales. Non--hit detectors retain a fixed background color. The RGB channels are defined as
\[
R_d = \mathrm{Norm}(T_d), \qquad  
G_d = \mathrm{Norm}(Q_d), \qquad  
B_d = |\,R_d - G_d\,| .
\]
The red channel emphasizes gradients of photon arrival times, the green channel spotlights relative brightness, while the blue channel encodes their contrast. In Figure \ref{fig: main_figure_1}, panel a and b, we show two examples of events transformed to our fingeprint. Note that the pixels corresponding to detectors without registered pulses are white.

This blue "opponent" channel $B_d$ highlights boundaries where timing and light yield diverge, which often occur near the edges of the Cherenkov light cone and thus carry strong directional information. The idea mirrors opponent representations in computer vision, where color--opponent and difference channels have long been used to decorrelate inputs and emphasize edges~\cite{rafegas2017color,simonyan2014twostream}. Our opponent channel accentuates directional cues that may otherwise be diluted in the raw features. The resulting \emph{fingerprints} compress thousands of raw pulses into compact, three--channel images that retain the spatiotemporal structure while being directly compatible with standard convolutional architectures.

\subsection{Network architecture}
We use a standard ResNet-18 architecture \cite{he2016deep} for image recognition as it has proven to be a competitive architecture in image recognition. The total number of parameters for this architecture is 11.188.074 which is comfortable lower than the total number of training examples. Since we are working with a regression task (we aim to predict the zenith and azimuth angle) we implement a final layer with only two nodes which output the predicted angles. Our activation function was selected through trial-and-error and we found that GeLU \cite{hendrycks2016gaussian} (out of ReLU, SiLU \cite{elfwing2018sigmoid} and ELU \cite{clevert2015fast}) provided the best performance for our task.

\subsection{Loss function, optimisation and data split}
We cast direction reconstruction as regression on the unit sphere $\mathbb{S}^2$. A direction parameterised by azimuth $\phi$ and zenith $\theta$ is mapped to a unit vector $\mathbf{u}(\alpha,\phi) = \big(\sin\phi\cos\alpha,\, \sin\phi\sin\alpha,\, \cos\phi\big)$.
We minimise the thee losses shown in Table \ref{tab:losses}. In all experiments we used Adam with learning rate $5\times 10^{-5}$, batch size $500$, and dropout $0.1$ which were selected by a small manual sweep. 

Since $\arccos$ has a derivative which diverges as $x\!\to\!\pm1$, we confine the cosine to $[-1+\varepsilon,\,1-\varepsilon]$ via $\mathrm{clamp}(x)=\min\!\big(\max(x,-1+\varepsilon),\,1-\varepsilon\big)$, with $\varepsilon=10^{-6}$. The Clamped Angular loss instead uses a straight-through clamp $\widetilde{\mathrm{clamp}}$ ($\varepsilon=10^{-5}$): its forward value equals $\mathrm{clamp}$, but within the clamped band it passes a unit gradient ($\partial\widetilde{\mathrm{clamp}}/\partial x=1$) rather than the zero gradient of an ordinary clamp.
\begin{table}[ht]
    \centering
    \caption{Loss functions used. All losses operate on predicted and true pairs $(\alpha, \phi)$, with $\cos\theta =
  \sin\phi_\mathrm{true}\sin\phi_\mathrm{pred}\cos(\alpha_\mathrm{true}-\alpha_\mathrm{pred}) + \cos\phi_\mathrm{true}\cos\phi_\mathrm{pred}$.}
    \label{tab:losses}
    \begin{tabular}{@{}lcc@{}}
    \toprule
    \textbf{Loss} & \textbf{Formula} & \textbf{Note} \\
    \midrule
    Angular
      & $\arccos\bigl[\mathrm{clamp}(\cos\theta_i)\bigr]$
      & \\[4pt]
    Clamped Angular
      & $\arccos\bigl[\widetilde{\mathrm{clamp}}(\cos\theta_i)\bigr]$
      &  \\[4pt]
    Combined Angular--vMF
      & $\arccos\bigl[\mathrm{clamp}(\cos\theta_i)\bigr]+\lambda\,\mathcal{L}_{\mathrm{vMF}}^{(i)}$
      & $\lambda=0.05$,\; $\kappa$ trainable \\
    \bottomrule
\end{tabular}
\vspace{4pt}
    {\small $\mathcal{L}_{\mathrm{vMF}}^{(i)}=-\kappa\cos\theta_i-\log C_3(\kappa)$,\quad
    $\log C_3(\kappa)=\log\kappa-\log\sinh\kappa-\log 4\pi$.}
\end{table}
\paragraph{Evaluation metric.}
Performance is reported as the mean angular error (MAE), i.e.\ the average geodesic distance between predicted and true directions,
\begin{equation*}
\label{eq:mae}
\mathrm{MAE}
= \frac{1}{N}\sum_{i=1}^{N} \arccos\!\Big(\mathbf{u}(\hat\alpha_i,\hat\phi_i)\cdot \mathbf{u}(\alpha_i,\phi_i)\Big).
\end{equation*}

\textbf{Dataset split.}  For training, validation and testing we use 660 (which corresponds to the total dataset) batches of samples. For testing and validation we consistently use batch 1 and batch 2 respectively, and use the remaining $n-2$ batches for training.

\newpage
\section{Results, Framing, and Outlook}
Our single, compact CNN---trained without ensembling or advanced optimisation (no LR schedules, cosine restarts, or augmentations)---achieves an MAE of $1.10\,\mathrm{rad}$ on the Kaggle dataset. 
For context, the top entries in the public \emph{IceCube--Neutrinos in Deep Ice} Kaggle competition, which combined graph neural networks and transformer modules with heavy ensembling, reached $\approx 0.95\,\mathrm{rad}$ MAE on the held-out leaderboard~\cite{bukhari2024icecube}. 
While our fingerprint CNN therefore falls short of absolute SOTA by about 
$0.15\,\mathrm{rad}$, it provides a clean convolutional baseline on the 
Kaggle dataset, where most high-ranking solutions have relied on 
graph neural networks or transformers.  Our result shows that with an appropriate event encoding, a straightforward CNN can still come within $\sim0.15\,\mathrm{rad}$ of the 
best competition entries, offering an interpretable and efficient baseline.

This work advances a methodological claim: a deterministic event--to--RGB encoding, where detector timing, brightness, and their discrepancy are mapped into image channel, yields effective direction reconstruction with conventional convolutional networks. The representation is intrinsically interpretable as each detector corresponds to a single pixel. This allows for straightforward diagnostic of network predictions. 

There is ample room for improvement without abandoning the simplicity: (i) better encodings that respect locality (e.g., space-filling curves \cite{verbruggen2023improving}, string$\times$depth rasters, additional color channels), (ii) modest architectural upgrades (depthwise-separable/modern CNN blocks) and schedule tuning, and (iii) alternative loss formulations based on the von~Mises--Fisher distribution, which extend the cosine objective by learning a concentration parameter and thereby provide calibrated directional uncertainties. A future ViT baseline on these fingerprints would test whether global attention adds value beyond convolution.

In conclusion, CNNs are not obsolete for IceCube-like reconstruction tasks: the proposed neutrino-fingerprint encoding provides an interpretable and efficient baseline that establishes a solid reference while leaving room for future advances in model and encoding design.
\begin{ack}
Brecht Verbeken acknowledges support from ENRICH (European Network for radiation-detection based Research and Innovation addressing increasing societal CHallenges) through COST Action CA24131. Vincent Ginis acknowledges support from the Research Foundation -- Flanders (FWO) under grants No.~G032822N and G0K9322N.
\end{ack}
\newpage
\bibliographystyle{unsrtnat}  
\bibliography{ref.bib}

\end{document}